\documentclass[apjl]{emulateapj}
\usepackage{apjfonts}

\newcommand{\na}{New Astronomy}

\begin{document}
\shorttitle{6 \lowercase{cm} Lines of OH in OH/IR Stars}
\shortauthors{Fish et al.}
\title{Observations of the 6 Centimeter Lines of OH in Evolved (OH/IR)
Stars}
\journalinfo{Submitted to ApJL on 21 September 2006, accepted 24
  October 2006}
\author{Vincent L.\ Fish\altaffilmark{1,2},
        Laura K.\ Zschaechner\altaffilmark{1,3},
        Lor\'{a}nt O.\ Sjouwerman\altaffilmark{1},
        Ylva M.\ Pihlstr\"{o}m\altaffilmark{4},
        Mark J.\ Claussen\altaffilmark{1}}
\altaffiltext{1}{National Radio Astronomy Observatory, 1003 Lopezville
Rd., Socorro, NM  87801}
\altaffiltext{2}{Jansky Fellow}
\altaffiltext{3}{Present address: Department of Physics and
    Astronomy, University of Montana, 32 Campus Dr.\ \#1080 Missoula,
    MT 59812}
\altaffiltext{4}{Department of Physics and Astronomy, University of
  New Mexico, 800 Yale Blvd.\ NE, Albuquerque, NM 87131}
\email{vfish@nrao.edu}

\begin{abstract}
Recent observational and theoretical advances have called into
question traditional OH maser pumping models in evolved (OH/IR) stars.
The detection of excited-state OH lines would provide additional
constraints to discriminate amongst these theoretical models.  In this
Letter, we report on VLA observations of the 4750 MHz and 4765 MHz
lines of OH toward 45 sources, mostly evolved stars.  We detect 4765 MHz
emission in the star forming regions Mon R2 and LDN 1084, but we do
not detect excited-state emission in any evolved stars.  The flux
density and velocity of the 4765~MHz detection in Mon~R2 suggests that
a new flaring event has begun.
\end{abstract}

\section{Introduction}

The standard model of the pumping mechanism behind 1612 MHz masers in
OH/IR stars was developed by \citet{Elitzur76}.  The masers are
believed to be pumped by 35 $\mu$m radiation, which connects the
ground states with the $^2\Pi_{1/2}, J = 5/2$ states.  Subsequent
decays down the $^2\Pi_{1/2}$ ladder and back to the ground states
invert the 1612 MHz transition when the infrared transitions are
optically thick.  Pumping by 53 $\mu$m radiation (via the
$^2\Pi_{1/2}, J = 3/2$ states) can also invert the 1612 MHz transition
by this mechanism \citep{Elitzur81}.

More recent detailed modelling by \citet{Gray05} offers greater
insight into 1612 MHz inversion.  The two primary ways of pumping the
1612 MHz maser are through collisional excitation within the
$^2\Pi_{3/2}$ ladder and radiative excitation by 53 $\mu$m photons via
the $^2\Pi_{1/2}, J = 3/2$ states, with the latter being the dominant
source of the inversion.  In the case of a large population transfer
through the $^2\Pi_{1/2}, J = 1/2, F = 1^-$ level, it is possible that
the 4750 MHz ($F = 1^- \rightarrow 1^+$) and 4765 MHz ($F = 1^-
\rightarrow 0^+$) $\Lambda$-doubling transitions will be seen in
addition to the 1612 MHz transition.

An important detail in these models is whether the radiative pump
operates primarily through the 35 $\mu$m infrared lines or the 53
$\mu$m lines.  The former are more energetic and therefore require a
larger far infrared radiation field to produce an efficient maser
pump.  Detailed hot envelope models can include a large contribution
from 35 $\mu$m radiation (M.~D.~Gray 2006, private communication).
Unfortunately, infrared data are inconclusive as to the relative
importance of these pumping routes.  An analysis of archive data from
the Infrared Space Observatory by \citet{He05} produced three
detections of 35~$\mu$m absorption in red supergiants, but also 15
nondetections among stellar objects where the absorption line should
be above the threshold of detectability, assuming the
\citet{Elitzur76} pump rates.  All sources with detected 35~$\mu$m
absorption also display 53~$\mu$m absorption with similar equivalent
widths \citep{He05}.  The authors suggest that 53~$\mu$m absorption
may play a role in pumping 1612 MHz masers in evolved stars,
consistent with \citet{Gray05}.

Detections of OH lines beyond the ground-state transitions could
provide useful constraints to clarify the modelling picture.  There
have been several searches for maser emission in the 4.7 GHz
($^2\Pi_{1/2}, J = 1/2$) and 6.0 GHz ($^2\Pi_{3/2}, J = 5/2$)
transitions of OH
\citep{Thacker70,Zuckerman72,Baudry74,Rickard75,Claussen81,Jewell85,Desmurs02}.
All searches failed to detect excited-state emission from OH/IR stars,
with the exception of 6035 MHz emission in NML Cyg \citep{Zuckerman72}
and 4750 MHz emission in AU Gem \citep{Claussen81}, neither of which
was confirmed in subsequent observations by \citet{Jewell85}.  But the
total number of observed OH/IR stars in the 4.7 GHz lines remains less
than three dozen.  Additionally, observers have generally chosen
sources with very bright far infrared fluxes, which may include an
inherent bias if negative optical depths in the 4.7 GHz lines are not
independent of the ratio of 35~$\mu$m to 53~$\mu$m fluxes.

Given the renewed interest in OH/IR star pumping generated by the
\citet{He04} and \citet{He05} observations as well as the
\citet{Gray05} theory, further searches for excited-state OH in
evolved stars are justified.  In this Letter, we present observations
of a larger sample of OH/IR stars and other sources in the 4750 and
4765~MHz lines of OH.

\section{Observations}

Data were taken during three sessions during 2006 May 25--27 using the
Very Large Array (VLA).  The observations occurred near the end of
reconfiguration between A and BnA configurations.  Several antennas
were out of the array due to the move as well as the EVLA upgrade,
leaving 22 antennas in operation.

The observed sources are drawn primarily from the \citet{Chen01}
catalogue of OH/IR sources and were selected based on their range of
observability in LST.  Several of these sources are not actually
evolved stars, likely due to incorrect IRAS associations at the
1\arcmin\ level.  The total time on each source was approximately
three minutes, with one-minute calibrator scans interspersed between
sources.  A significant amount of radio frequency interference (RFI)
was observed on two sources, IRAS 05437$-$0001 and IRAS 06053$-$0622,
although simple flagging of the affected time ranges resulted in
sufficient data apparently free of RFI to produce reasonable images.

The 4750.656 and 4765.562 MHz transitions of OH were observed
simultaneously in both circular polarizations.  The 1.5625 MHz
bandwidth was centered at the LSR velocity of each source listed in
Table \ref{tab-results} and divided into 128 spectral channels, giving
a channel spacing of 12.207 kHz ($0.77$~km\,s$^{-1}$).

Data reduction was performed in the Astronomical Image Processing
System \citep[AIPS, ][]{Greisen03}.  Image cubes of the central
$80$~km\,s$^{-1}$ and measuring 10\arcmin\ in Right Ascension and
Declination were created using IMAGR.  Each channel was visually
scanned for maser emission and analyzed for the maximum pixel value
and rms noise.  Detections and upper limits for nondetections are
listed in Table \ref{tab-results}.
     
The snapshot observations provided a single-channel rms noise of $\sim
20$~mJy, which would allow for a $5\,\sigma$ detection of a 100~mJy
maser source.  This is the flux density of the lone detection of a 4.7
GHz maser in the Mira variable AU Gem \citep{Claussen81}.  The
distance of this star, 2.4~kpc \citep{Nguyen-Q-Rieu79}, is typical for
our sample.

\section{Results}

\subsection{Detections}
\label{Detections}

There were no detections among the OH/IR stars, but two previously
known 4765 MHz masers were observed in star forming regions.  The
first of these sources, IRAS 21413+5442 (LDN 1084), is a region of
massive star formation affiliated with an ultracompact \ion{H}{2}
region \citep{Cohen88}.  From our observations, the detected maser has
a flux density of 250 mJy in the channel centered at
$-61.80$~km\,s$^{-1}$ and 260 mJy in the channel centered at
$-62.57$~km\,s$^{-1}$.  The data are consistent with a single point
source at a velocity near $-62.2$~km\,s$^{-1}$ and located at
$21^\mathrm{h}43^\mathrm{m}01\fs452, +54\degr56\arcmin17\farcs87$
(J2000).  This agrees with the position of the $-62.10$~km\,s$^{-1}$
feature detected by \citet{Harvey-Smith05} to within $0\farcs25$.
  The second detected source is IRAS 06053$-$0622, (Mon R2).  Its flux
density is approximately 2.5~Jy in the channel centered at
$10.4$~km\,s$^{-1}$, located at $06^\mathrm{h}07^\mathrm{m}47\fs845,
-06\degr22\arcmin56\farcs61$, which agrees to within $0\farcs08$ with
the position of the brighter 10.62~km\,s$^{-1}$ maser detected by
\citet{Harvey-Smith05}.

Because the width of the channels used during the observations is
larger than a typical maser width in a star forming region, the quoted
flux densities are lower limits.  A closer approximation can be found
by dividing the channel width by the estimated velocity range of the
maser and then multiplying by the observed flux density.  Assuming a
single maser whose linewidth is 0.4~km\,s$^{-1}$ \citep[an average
value for Mon R2; see][]{Smits98} yields flux densities of just under
1 Jy for LDN 1084 and 5 Jy for Mon R2.

\subsection{Variability}

As is common for masers, both LDN 1084 and Mon R2 have displayed a
certain degree of variability in the past.  The flux density of the
4765 MHz maser in LDN 1084 was observed to be 700 mJy in 1989 and
again in 1991 \citep{Cohen91,Cohen95} but had dropped to 480 mJy by
1995 \citep{Harvey-Smith05}.  Our data suggest that the maser flux
density has since increased.  The velocity of this feature,
$-62.1$~km\,s$^{-1}$, is in the middle of the range of 6035 MHz
emission \citep{Fish06} and is consistent with our measurements given
our coarse spectral resolution.

In Mon R2 4765~MHz emission was first detected at 10.9~km\,s$^{-1}$ by
\citet{Gardner83}.  \citet{Cohen95} confirmed its status as a maser
and noticed variability, with a peak flux of 1.5 Jy in 1990.
Subsequent monitoring caught two flares to a maximum of nearly
$80$~Jy, with the central maser velocity varying between about 10.55
and 10.85~km\,s$^{-1}$ \citep{Smits98,Smits03}.  In 2000,
\citet{Dodson02} failed to detect any 4765~MHz emission in Mon R2 at
the 80~mJy level, and \citet{Smits03} found no 4765~MHz emission in
Mon R2 between 1998 December and the end of their observations in 2001
November despite monitoring the source at two-week intervals.

It appears that emission from the 4765 MHz maser(s) in Mon R2 has
returned.  The LSR velocity of our detection is consistent with being
near or just below the low end of the aforementioned velocity range.
We detect strong emission in the channel centered at
10.44~km\,s$^{-1}$ and possible weak emission in the next-lower
velocity channel (9.67~km\,s$^{-1}$), but not in the next-higher
channel (11.20~km\,s$^{-1}$).

\section{Discussion}

We do not detect 4.7 GHz OH maser emission from any of the evolved
stars in our sample.  This is consistent with most previous surveys of
evolved stars, in which no excited-state emission is detected
\citep{Thacker70,Baudry74,Rickard75,Jewell85,Desmurs02}.
Nevertheless, excited-state emission \emph{has} been detected in two
evolved stars.  \citet{Zuckerman72} report on 6035 MHz (and possibly
6030 MHz) maser emission in the red supergiant NML Cyg, although not
at the same velocity as the 1612 MHz masers.  Likewise,
\citet{Claussen81} report on 4750 MHz emission from the Mira AU Gem;
again the velocities are not the same as in ground-state emission,
although the authors note that the spectrum of the 4750 MHz emission
appears to be centered at the same velocity, with peaks nearer the
central velocity than at 1667 MHz.  Both of these appeared to be
convincing detections, yet the 6035 MHz emission in NML Cyg has
disappeared \citep{Jewell85,Desmurs02}.  The 4750 MHz emission in AU
Gem was also not redetected in observations by \citet{Jewell85},
although the 100~mJy maser would only have been 1.5 times their rms
noise, which does not conclusively establish that the maser had
disppeared.  Nevertheless, it appears that excited-state emission in
late-type stars is both rare and time-variable.

There are several classes of theoretical models for OH pumping in
circumstellar shells.  The \citet{Elitzur76} model of pumping via the
35~$\mu$m lines and subsequent decay down the $^2\Pi_{1/2}$ ladder
inverts the 1612 MHz line.
Pumping via the less-energetic 53~$\mu$m lines can also produce a
strong 1612 MHz inversion \citep{Elitzur81,Gray05}.  Collisions may
contribute a substantial fraction of the inversion \citep{Gray05}.
Far infrared line overlaps can invert the 1612, 1665, and 1667 MHz
lines because of asymmetries in the dipole matrix elements
\citep{Bujarrabal80a,Bujarrabal80b}.  Near infrared line overlap,
possibly with H$_2$O, also may be necessary to account for main-line
maser emission \citep{Cimerman80,Collison93,Collison94}.

A combination of several of these pumps may occur in evolved stars,
either in the same spatial region or at different radii.
Observational evidence supports several of these pumping models.
Multiple infrared lines of OH have been detected
\citep{Sylvester97,He04,He05}.  Interferometric measurements indicate
that main-line masers exist at smaller radii than 1612 MHz masers in
circumstellar envelopes \citep{Harvey74}.  Substantial qualitative
differences in variability confirm this \citep{Etoka00} and likely
indicate that dust reprocessing of radiation is a critical element of
the radiative pumping \citep{Elitzur78}.  The spectrum of the lone
4750 MHz detection in AU Gem also suggests that the right conditions
for masing in excited-state lines exist interior to the region
producing main-line masers \citep{Claussen81}.  This is analogous to
H$_2$O and 1612 MHz OH maser observations in other stars, in which the
H$_2$O masers are typically seen at smaller expansion velocities than
the OH masers and are observed to exist at smaller radii as well
\citep{Habing96}.

Why, then, are excited-state masers in late-type stars so rare?  Three
ingredients are essential to produce a detectable OH maser: a
sufficient column density of OH, velocity coherence, and effective
pumping conditions.  The abundance of OH masers in the envelopes of
evolved stars clearly shows that column density of OH is sufficient to
produce maser activity.  But velocity coherence may be a problem in
the excited states.  It appears that 4750 MHz masers are located at
smaller radii than 1612, 1665, and 1667 MHz masers.  It is possible
that the velocity field at this radius is too irregular to support
large coherent path lengths, due possibly to accelerations or
turbulence.

It is also possible that the conditions necessary to pump
excited-state masers are very fragile.  Many pairs of far infrared OH
transitions overlap at different Doppler shifts on the order of
several km\,s$^{-1}$ \citep[see Table 2 of][]{Collison93}.  Small
changes in the velocity structure of a circumstellar envelope may
therefore have large consequences in the pumping.  When combined with
other pump mechanisms, such as collisional excitation, these effects
may be exaggerated.  Detailed modelling including the effects of
multiple pump mechanisms may be required to discover the precise
physical conditions responsible for producing detectable excited-state
maser emission.

More observations will be required in order to understand the narrow
range of parameter space conducive to excited-state OH pumping.  In
particular, the two late-type stars in which excited-state maser
emission was detected in one epoch (NML Cyg at 6035 MHz and AU Gem at
4750 MHz) should be reobserved with greater sensitivity.  If they are
redetected, their variability and spatial distribution will provide
important clues as to the pumping mechanism responsible for
excited-state maser emission in evolved stars.

\acknowledgments

The National Radio Astronomy Observatory is a facility of the National
Science Foundation (NSF) operated under cooperative agreement by
Associated Universities, Inc.  L.~K.~Z.\ acknowledges support from the
NSF Research Experiences for Undergraduates (REU) program.  This
research has made use of the SIMBAD database, operated at CDS,
Strasbourg, France.

\textit{Facility: VLA}

\clearpage

\begin{deluxetable}{llrlllrrrrr}
\setlength{\tabcolsep}{0.03in}
\tabletypesize{\footnotesize}
\tablecaption{Results of Observations at 4750 and 4765 MHz \label{tab-results}}
\tablehead{
%Row 0
   \colhead {IRAS} &
   \colhead {RA\tablenotemark{a}} & 
   \colhead {Dec\tablenotemark{a}} &
   \colhead {Source} &
   \colhead {Name} &
   \colhead {Alias} &
   \colhead {RAFGL} &
   \colhead {v$_\mathrm{LSR}$\tablenotemark{a}} &
   \multicolumn {2} {c} {Flux Density\tablenotemark{c}} &
   \colhead{Source}\\
%Row 1
   \colhead {Name}&
   \colhead {}&
   \colhead {}&
   \colhead {Type\tablenotemark{b}}&
   \colhead {}&
   \colhead {}&
   \colhead {Name}&
   \colhead {}&
   \colhead{4750}&
   \colhead{4765}& 
   \colhead{Type}\\
%Row 2
   \colhead {}&
   \multicolumn {2} {c} {(J2000)} &
   \colhead {} &
   \colhead {} &
   \colhead {} &
   \colhead {} &
   \colhead {(km\,s$^{-1}$)} &
   \multicolumn {2} {c} {(mJy)}&
   \colhead {Refs}                 
}
\startdata
{00170+6542}   & 00:19:51.5  & +65:59:31    & OH/IR & \nodata  & OH 119.7+3.3&     & $-51.3$  & $<$77 & $<$82 &  1 \\
{00428+6854}   & 00:46:00.6  & +69:10:54    & Mira  & V524 Cas & IRC +70012 &  107 & $-25.5$  & $<$76 & $<$80 &  2 \\
{01037+1219}   & 01:06:25.9  & +12:35:54    & OH/IR & WX Psc   & IRC +10011 &  157 &   $8.8$  & $<$81 & $<$86 &  3 \\
{01085+3022}   & 01:11:15.9  & +30:38:05    & Mira  & AW Psc   & IRC +30021 &  168 & $-26.8$  & $<$84 & $<$93 &  3 \\
{01110+2652}   & 01:13:47.8  & +27:07:56    & SR    & RT Psc   & IRC +30023 &  179 &$-116.4$  & $<$90 & $<$94 &  4 \\
{01304+6211}   & 01:33:50.6  & +62:26:47    & Mira  & V669 Cas & OH 127.8+0.0& 230 & $-55.0$  & $<$78 & $<$82 &  5 \\
{01572+5844}   & 02:00:44.1  & +58:59:03    & OH/IR & \nodata  &            &      & $-11.4$  & $<$77 & $<$82 &  6 \\
{02192+5821}   & 02:21:51.1  & +58:35:08    & OH/SG & S Per    & IRC +60088 &  323 & $-38.5$  & $<$83 & $<$87 &  7 \\
{02420+1206}   & 02:44:45.0  & +12:19:00    & Mira  & RU Ari   &            &       & $19.8$  & $<$82 & $<$86 &  2 \\
{02547+1106}   & 02:57:27.2  & +11:18:04    & OH/IR & YZ Ari   &            & 5087  & $15.6$  & $<$82 & $<$86 &  3 \\
{03206+6521}   & 03:25:08.5  & +65:32:05    & OH/IR & \nodata  & OH 138.0+7.2&     & $-37.5$  &$<$104 & $<$76 &  6 \\
{03293+6010}   & 03:33:30.5  & +60:20:09    & OH/IR & \nodata  & OH 141.7+3.5&5097 & $-57.5$  & $<$77 & $<$82 &  1 \\
{03507+1115}   & 03:53:28.6  & +11:24:20    & Mira  & IK Tau   & IRC +10050 &  529  & $33.9$  & $<$82 &$<$106 &  8 \\
{04130+3918}   & 04:16:24.6  & +39:25:44    & Carbon& C* 192   &            & 6312  & $-5.7$  & $<$72 & $<$91 &  9 \\
{04396+0647}   & 04:42:21.5  & +06:52:39    & OH/IR & BZ Tau   & IRC +10068 &  619  & $18.2$  & $<$81 &$<$102 &  3 \\
{04505$-$1006} & 04:52:57.7  & $-$10:02:00  & OH/IR & EY Eri   &            &       & $17.1$  & $<$77 & $<$98 & 10 \\
{04575+1251}   & 05:00:23.9  & +12:56:06    & OH/IR & \nodata&OH 187.7$-$17.6& 5134 & $0.9$  & $<$76 &$<$100 &  1 \\
{05274$+$3345} & 05:30:45.6  & +33:47:52    & SFR   & \nodata  &            & 5142  & $-4.1$  & $<$70 & $<$74 & 11 \\
{05358$-$0704} & 05:38:18.8  & $-$07:02:27  & FU Ori& V883 Ori &            &4433S   & $5.3$  & $<$84 &$<$103 & 12 \\
{05373$-$0810} & 05:39:42.6  & $-$08:09:08  & Carbon& V1187 Ori& IRC $-$10095& 796  & $11.2$  & $<$81 &$<$103 & 13 \\
{05380$-$0728} & 05:40:27.7  & $-$07:27:28  & FU Ori& Reipurth 50& HBC 494  & 5163   & $3.9$  & $<$81 &$<$102 & 12 \\
{05423+2905}   & 05:45:29.7  & +29:07:04    & OH/IR & V530 Aur &            &       & $28.7$  & $<$77 & $<$84 &  6 \\
{05437$-$0001} & 05:46:17.8  & $-$00:00:17  & SFR   & LDN 1627 & M 78       &       & $13.1$  &$<$100 &$<$130 & 14 \\
{06053$-$0622} & 06:07:46.7  & $-$06:23:00  & SFR   & Mon R2   &            &  877  & $8.9$  &$<$120 &$2500$\tablenotemark{d} & 15 \\
{06297+4045}   & 06:33:14.9  & +40:42:50    & OH/IR & \nodata  & IRC +40156 &  955 & $-16.0$  & $<$77 & $<$83 & 10 \\ 
{06319+0415}   & 06:34:37.6  & +04:12:44    & (P)PN & Rosette  &            &  961  & $12.7$  & $<$90 &$<$111 &  5 \\
{06500+0829}   & 06:52:46.9  & +08:25:20    & OH/IR & GX Mon   & IRC +10143 & 1028 & $-10.7$  & $<$70 & $<$93 & 16 \\
{07209$-$2540} & 07:22:59.2  & $-$25:46:08  & OH/SG & VY CMa   & IRC $-$30087&1111  & $22.4$  & $<$75 & $<$83 &  7 \\
{07331+0021}   & 07:35:41.2  & +00:14:59    & (P)PN & AI CMi   &            & 5236  & $28.4$  & $<$76 & $<$83 & 17 \\
{07399$-$1435} & 07:42:17.1  & $-$14:42:50  & (P)PN & QX Pup   &OH 231.8+4.2& 5237  & $15.9$  & $<$76 & $<$81 & 18 \\
{07445$-$2613} & 07:46:37.8  & $-$26:20:34  & Mira  & SS Pup  &OH 242.4$-$0.7&1192   & $82.7$ & $<$76 & $<$82 & 16 \\
{07585$-$1242} & 08:00:50.6  & $-$12:50:31  & Mira  & U Pup    & IRC $-$10184&1215   &$-15.6$ & $<$75 & $<$81 & 16 \\
{08005$-$2356} & 08:02:40.6  & $-$24:04:43  & (P)PN & V510 Pup &            &        & $-0.1$ & $<$87 & $<$75 & 19 \\
{08357$-$1013} & 08:38:08.8  & $-$10:24:17  & OH/IR & \nodata &OH 235.3+18.1& 1274   & $17.8$ & $<$72 & $<$76 & 10 \\
{09429$-$2148} & 09:45:17.0  & $-$22:01:56  & OH/IR & IW Hya   & IRC $-$20197&5259   & $39.0$ & $<$75 & $<$78 & 10 \\
{21413+5442}   & 21:43:01.4  & +54:56:16    & SFR   & LDN 1084 &            &       & $-61.8$ & $<$83 & $260$\tablenotemark{d} & 20 \\
{21554+6204}   & 21:56:58.3  & +62:18:43    & OH/IR & GLMP 1048&            &       & $-21.4$ & $<$75 & $<$77 &  6 \\
{22176+6303}   & 22:19:18.2  & +63:18:46    & SFR   & S140     &            & 2884   & $-7.0$ & $<$96 & $<$81 &  8 \\
{22177+5936}   & 22:19:27.9  & +59:51:21    & OH/IR & \nodata  &OH 104.9+2.5& 2885  & $-24.9$ & $<$76 & $<$77 &  1 \\
{22466+6942}   & 22:48:14.2  & +69:58:29    & OH/IR & V708 Cep &            &       & $-45.9$ & $<$79 & $<$81 &  6 \\
{22525+6033}   & 22:54:32.0  & +60:49:38    & OH/SG & MY Cep   &            & 2987   & $-0.8$ & $<$79 & $<$81 &  7 \\
{22556+5833}   & 22:57:41.3  & +58:49:15    & Symb  & V627 Cas & HBC 316    & 2999  & $-49.5$ & $<$74 & $<$77 & 21 \\
{23352+5834}   & 23:37:40.1  & +58:50:47    & Mira  & V850 Cas &            &       & $-0.1$  & $<$77 & $<$92 & 22 \\
{23416+6130}   & 23:44:03.6  & +61:47:22    & OH/SG & PZ Cas   & IRC +60417 & 3138 & $-38.5$  & $<$63 & $<$65 &  7 \\
{23425+4338}   & 23:45:02.1  & +43:55:22    & OH/IR & EY And   & IRC +40545 & 3143 & $-42.5$  & $<$84 & $<$88 & 10 \\
\enddata

\tablenotetext{a}{Pointing center and central LSR velocity of
observations, taken from the \citet{Chen01} catalogue.}
\tablenotetext{b}{Carbon: Carbon Star;
                  OH/SG: OH/Supergiant; 
                  (P)PN: (Proto)-Planetary Nebula/post-AGB star;
                  SFR: H~\textsc{ii} Region/Star Forming Region;
                  SR: Semiregular Variable;
                  Symb: Symbiotic.
                  Note that the status of several sources is disputed
                  and that not all categories are mutually exclusive.}
\tablenotetext{c}{Upper limits are $4\,\sigma$.}

\tablenotetext{d}{Detected flux density in a single channel of
                  effective resolution 0.92~km\,s$^{-1}$.  See \S
                  \ref{Detections} for a discussion of likely maser
                  width and peak flux density.  The maser in Mon~R2
                  was detected in the channel centered at
                  $10.4$~km\,s$^{-1}$.}

\tablerefs{(1) \citet{LeSqueren92};  (2) \citet{Szymczak99};
           (3) \citet{Lewis00};      (4) \citet{Jura92};
           (5) \citet{He05};         (6) \citet{Galt89};
           (7) \citet{Lewis91};      (8) \citet{He04};
           (9) \citet{Cohen96};      (10) \citet{Lewis01};
           (11) \citet{Snell88};     (12) \citet{Strom93};
           (13) \citet{Szczerba02};  (14) \citet{Parker91};
           (15) \citet{Knapp76};     (16) \citet{Sivagnanam88};
           (17) \citet{Omont93};     (18) \citet{teLintelHekkert88};
           (19) \citet{Slijkhuis91}; (20) \citet{Cohen88};
           (21) \citet{Kolotilov96}; (22) \citet{Sivagnanam90}}
\end{deluxetable}


\begin{thebibliography} {}

\bibitem[Baudry(1974)]{Baudry74} Baudry, A.\ 1974, \aap, 32, 191

\bibitem[Bujarrabal et al.(1980a)]{Bujarrabal80a} Bujarrabal, V.,
  Destombes, J.~L., Guibert, J., Marli\`{e}re-Demuynck, C.,
  Nguyen-Q-Rieu, \& Omont, A.\ 1980a, \aap, 81, 1

\bibitem[Bujarrabal et al.(1980b)]{Bujarrabal80b} Bujarrabal, V.,
  Guibert, J., Nguyen-Q-Rieu, \& Omont, A.\ 1980b, \aap, 84, 311

\bibitem[Chen et al.(2001)]{Chen01} Chen, P.~S., Szczerba, R., Kwok,
  S., \& Volk, K.\ 2001, \apj, 368, 1006

\bibitem[Cimerman \& Scoville(1980)]{Cimerman80} Cimerman, M., \&
  Scoville, N.\ 1980, \apj, 239, 526

\bibitem[Claussen \& Fix(1981)]{Claussen81} Claussen, M.~J., \& Fix,
  J.~D.\ 1981, \apjl, 250, L77

\bibitem[Cohen et al.(1988)]{Cohen88} Cohen, R.~J., Baart, E.~E., \&
  Jonas, J.~L.\ 1988, \mnras, 231, 205

\bibitem[Cohen et al.(1995)]{Cohen95} Cohen, R.~J., Masheder,
  M.~R.~W., \& Caswell, J.~L.\ 1995, \mnras, 274, 808

\bibitem[Cohen et al.(1991)]{Cohen91} Cohen, R.~J., Masheder,
  M.~R.~W., \& Walker, R.~N.~F.\ 1991, \mnras, 250, 611

\bibitem[Cohen et al.(1996)]{Cohen96} Cohen, M., Wainscoat, R.~J., 
  Walker, H.~J., \& Volk, K.\ 1996, \aj, 111, 1333

\bibitem[Collison \& Nedoluha(1993)]{Collison93} Collison, A.~J., \&
  Nedoluha, G.~E.\ 1993, \apj, 413, 735

\bibitem[Collison \& Nedoluha(1994)]{Collison94} Collison, A.~J., \&
  Nedoluha, G.~E.\ 1994, \apj, 422, 193

\bibitem[Desmurs et al.(2002)]{Desmurs02} Desmurs, J.-F., Baudry, A.,
  Sivaganam, P., \& Henkel, C.\ 2002, \aap, 394, 975

\bibitem[Dodson \& Ellingsen(2002)]{Dodson02} Dodson, R.~G., \&
  Ellingsen, S.~P.\ 2002, \mnras, 333, 307

\bibitem[Elitzur(1978)]{Elitzur78} Elitzur, M.\ 1978, \aap, 62, 305

\bibitem[Elitzur(1981)]{Elitzur81} Elitzur, M.\ 1981, ASSL 
  Vol.~88: Physical Processes in Red Giants, 363 

\bibitem[Elitzur et al.(1976)]{Elitzur76} Elitzur, M., Goldreich, P.,
  \& Scoville, N.\ 1976, \apj, 205, 384

\bibitem[Etoka \& Le Squeren(2000)]{Etoka00} Etoka, S., \& Le Squeren,
  A.~M.\ 2000, \aap{}S, 146, 179

\bibitem[Fish et al.(2006)]{Fish06} Fish, V.~L., Reid, M.~J., Menten,
  K.~M., \& Pillai, T.\ 2006, \aap, in press, astro-ph/0608121

\bibitem[Galt et al.(1989)]{Galt89} Galt, J.~A., Kwok, S., \& Frankow,
  J.\ 1989, \aj, 98, 2182

\bibitem[Gardner \& Mart\'{\i}n-Pintado(1983)]{Gardner83} Gardner,
  F.~F., \& Mart\'{\i}n-Pintado, J.\ 1983, \aap, 121, 265

\bibitem[Gray et al.(2005)]{Gray05} Gray, M.~D., Howe, D.~A., \&
  Lewis, B.~M.\ 2005, \mnras, 364, 783

\bibitem[Greisen(2003)]{Greisen03} Greisen, E.~W.\ 2003, AIPS, the
  VLA, and the VLBA in Information Handling in Astronomy --
  Historical Vistas, ed.\ A.\ Heck (Dordrecht: Kluwer Academic
  Publishers), 109

\bibitem[Habing(1996)]{Habing96} Habing, H.~J.\ 1996, \aapr, 7, 97

\bibitem[Harvey et al.(1974)]{Harvey74} Harvey, P.~J., Booth, R.~S.,
  Davies, R.~D., Whittet, D.~C.~B., \& McLaughlin, W.\ 1974, \mnras,
  169, 545

\bibitem[Harvey-Smith \& Cohen(2005)]{Harvey-Smith05} Harvey-Smith,
  L., \& Cohen R. J.\ 2005, \mnras, 365, 637

\bibitem[He \& Chen(2004)]{He04} He, J.~H., \& Chen, P.~S.\ 2004, \na,
  9, 545

\bibitem[He et al.(2005)]{He05} He, J.~H., Szczerba, R., Chen, P.~S.,
  \& Sobolev, A.~M.\ 2005, \aap, 434, 201

\bibitem[Jewell et al.(1985)]{Jewell85} Jewell, P.~R., Schenewerk,
  M.~S., \& Snyder, L.~ E.\ 1985, \apj, 295, 183

\bibitem[Jura \& Kleinmann(1992)]{Jura92} Jura, M., \& Kleinmann,
  S.~G.\ 1992, \apjs, 83, 329

\bibitem[Knapp \& Brown(1976)]{Knapp76} Knapp, G.~R., \& Brown, R.~L.\
  1976, \apj, 204, 21

\bibitem[Kolotilov et al.(1996)]{Kolotilov96} Kolotilov, E.~A.,
  Munari, U., Yudin, B.~F., \& Tatarnikov, A.~M.\ 1996, Astronomy
  Reports, 40, 812

\bibitem[Le Squeren et al.(1992)]{LeSqueren92} Le Squeren, A.~M.,
  Sivagnanam, P., Dennefeld, M., \& David, P.\ 1992, \aap, 254, 133

\bibitem[Lewis(1991)]{Lewis91} Lewis, B.~M.\ 1991, \aj, 101, 254

\bibitem[Lewis(2000)]{Lewis00} Lewis, B.~M.\ 2000, \apj, 533, 959

\bibitem[Lewis(2001)]{Lewis01} Lewis, B.~M.\ 2001, \apj, 560, 400

\bibitem[Nguyen-Q-Rieu et al.(1979)]{Nguyen-Q-Rieu79} Nguyen-Q-Rieu,
  Laury-Micoulaut, C., Winnberg, A., \& Schultz, G.~V.\ 1979, \aap,
  75, 351

\bibitem[Omont et al.(1993)]{Omont93} Omont, A., Loup, C., Forveille,
  T., te Lintel Hekkert, P., Habing, H., \& Sivagnanam, P.\ 1993,
  \aap, 267, 515

\bibitem[Parker(1991)]{Parker91} Parker, N.~D.\ 1991, \mnras, 251, 63

\bibitem[Rickard et al.(1975)]{Rickard75} Rickard, L.~J., Zuckerman,
  B., \& Palmer, P.\ 1975, \apj, 200, 6

\bibitem[Sivagnanam et al.(1988)]{Sivagnanam88} Sivagnanam, P., Le
  Squeren, A.~M., \& Foy, F.\ 1988, \aap, 206, 285

\bibitem[Sivagnanam et al.(1990)]{Sivagnanam90} Sivagnanam, P., Le
  Squeren, A.~M., Minh, F.~T., \& Braz, M.~A.\ 1990, \aap, 233, 112

\bibitem[Slijkhuis et al.(1991)]{Slijkhuis91} Slijkhuis, S., Hu.,
  J.~Y., \& de Jong, T.\ 1991, \aap, 248, 547

\bibitem[Sloan \& Price(1998)]{Sloan98} Sloan, G.~C., \& Price, S.~D.\
  1998, \apjs, 119, 141

\bibitem[Smits(2003)]{Smits03} Smits D.~P.\ 2003, \mnras, 339, 1

\bibitem[Smits et al.(1998)]{Smits98} Smits, D.~P., Cohen, R.~J., \&
  Hutawarakorn, B.\ 1998, \mnras, 296, L11

\bibitem[Snell et al.(1988)]{Snell88} Snell, R.~L., Huang, Y.-L.,
  Dickman, R.~L., \& Claussen, M.~J.\ 1988, \apj, 325, 853

\bibitem[Strom \& Strom(1993)]{Strom93} Strom, K.~M., \& Strom, S.~E.\
  1993, \apjl, 412, L63

\bibitem[Sylvester et al.(1997)]{Sylvester97} Sylvester, R.~J., et
  al.\ 1997, \mnras, 291, L42

\bibitem[Szczerba et al.(2002)]{Szczerba02} Szczerba, R., Chen, P.~S.,
  Szymczak, M., \& Omont, A.\ 2002, \aap, 381, 491

\bibitem[Szymczak \& Le Squeren(1999)]{Szymczak99} Szymczak, M., \& Le
  Squeren, A.~M.\ 1999, \mnras, 304, 415

\bibitem[te Lintel Hekkert et al.(1988)]{teLintelHekkert88} te Lintel
  Hekkert, P., Habing, H.~J., Caswell, J.~L., Norris, R.~P., \&
  Haynes, R.~F.\ 1988, \aap, 202, L19

\bibitem[Thacker et al.(1970)]{Thacker70} Thacker, D.~L., Wilson,
  W.~J., \& Barrett, A.~H.\ 1970, \apjl, 161, L191

\bibitem[Zuckerman et al.(1972)]{Zuckerman72} Zuckerman, B., Yen,
  J.~L., Gottlieb, C.~A., \& Palmer, P.\ 1972, \apj, 177, 59

\end{thebibliography}
\end{document}